\newcommand{\NPA}[3]{Nucl.\ Phys.\ A\ {\bf #1},\ #2 (#3)}

\newcommand{\PLB}[3]{Phys.\ Lett.\ B\ {\bf #1},\ #2 (#3)}

\newcommand{\PRL}[3]{Phys.\ Rev.\ Lett.\ {\bf #1},\ #2 (#3)}

\newcommand{\PRC}[3]{Phys.\ Rev.\ C\ {\bf #1},\ #2 (#3)}
\newcommand{\PRD}[3]{Phys.\ Rev.\ D\ {\bf #1},\ #2 (#3)}

\newcommand{\diracslash}[1]{#1\llap{/\kern2pt}}

\newcommand{\be}{\begin{equation}}
\newcommand{\ee}{\end{equation}}
\newcommand{\bea}{\begin{eqnarray}}
\newcommand{\eea}{\end{eqnarray}}
\newcommand{\ba}[1]{\begin{array}{#1}}
\newcommand{\ea}{\end{array}}

\documentclass[prd,aps,floats,nofootinbib,tightenlines,showpacs]{revtex4}
\usepackage{epsfig,graphicx,pstricks}
\usepackage{psfrag}
\usepackage{color}
\usepackage{amsmath}
\usepackage{amsfonts}
\usepackage{amssymb}
\usepackage{textcomp}
\usepackage{multirow}
\usepackage{subfigure}
\usepackage{slashed}

\begin{document}

\title {Medium modification of hadron masses and the thermodynamics of hadron resonance gas model}
\author{Guru Prakash Kadam }
\email{guruprasad@prl.res.in}
\affiliation{Theory Division, Physical Research Laboratory,
Navrangpura, Ahmedabad 380 009, India}
\author{Hiranmaya Mishra}
\email{hm@prl.res.in}
\affiliation{Theory Division, Physical Research Laboratory,
Navrangpura, Ahmedabad 380 009, India}

\date{\today} 

\def\be{\begin{equation}}
\def\ee{\end{equation}}
\def\bearr{\begin{eqnarray}}
\def\eearr{\end{eqnarray}}
\def\zbf#1{{\bf {#1}}}
\def\bfm#1{\mbox{\boldmath $#1$}}
\def\hf{\frac{1}{2}}
\def\sl{\hspace{-0.15cm}/}
\def\omit#1{_{\!\rlap{$\scriptscriptstyle \backslash$}
{\scriptscriptstyle #1}}}
\def\vec#1{\mathchoice
        {\mbox{\boldmath $#1$}}
        {\mbox{\boldmath $#1$}}
        {\mbox{\boldmath $\scriptstyle #1$}}
        {\mbox{\boldmath $\scriptscriptstyle #1$}}
}

\begin{abstract}
We study the effect of temperature (T) and baryon density ($\mu$) dependent hadron masses on the thermodynamics of hadronic matter.  We use linear scaling rule in terms of constituent quark masses for all hadrons except for light mesons. T and $\mu$ dependent constituent quark masses and the light mesons masses are computed using 2+1 flavor Nambu-Jona-Lasinio (NJL) model.  We compute the thermodynamical quantities of hadronic matter within excluded volume hadron resonance gas model (EHRG) with these T and $\mu$ dependent hadron masses. We confront the thermodynamical quantities with the lattice quantum chromodynamics (LQCD) at $\mu=0$ GeV. Further, we comment on the effect of T and $\mu$ dependent hadron masses on the transport properties near the transition temperature ($T_{c}$). 
\end{abstract}


\maketitle

\section{Introduction}

 The hadron resonance gas model (HRG) and its extended version, excluded volume HRG (EHRG) is the simplest and the most successful effective model of quantum chromodynamics (QCD) describing the hadronic phase of strongly interacting matter. The HRG model is essentially based on the result that the interacting hadron resonance matter can be approximated by that of non-interacting gas of hadrons and all the resonances\cite{Huovinen}. Before this, there was an interesting result obtained in Ref. \cite{welke,prakashvenu} where the authors showed that the interacting pion gas behaves similar to the ideal system containing pions and $\rho$ mesons. The HRG model has also been confronted with the lattice QCD (LQCD) and found to be in a very good agreement\cite{prasad}.  It has been extensively used in the literature for various purposes, $viz.$, to obtain hadron yield in heavy ion collision experiments\cite{cleymansatz,Becattini,magestro,rafelski}, estimating transport properties of hadronic matter like shear viscosity\cite{prakash,prakashwiranata,khvorostukhin,greinerprc,gorenstein,cpsingh,tawfik,greinerprl,gurunpa,gurumpla,guruprc} and the study of fluctuations in conserved charges in heavy ion collision experiments\cite{bhattacharyya,garg}. 
 
 Albeit non-interacting HRG is in agreement with LQCD, it misses one important feature of hadronic interactions; the repulsive interactions. The existence of repulsive interactions has been confirmed from nucleon-nucleon scattering experiments and, in fact, used to assign finite hard-core radius to nucleons\cite{heppe,bohr}. Further necessity to include short range repulsive interaction comes from heavy ion collision experiments. It has been observed that the chemical freeze-out parameters obtained from fitting the particle number ratios at AGS and SPS energies lead to large values of total particle number densities. One can suppress these number densities by incorporating the repulsive interactions in HRG model through excluded volume corrections\cite{ehrgrishke,ehrgclaymans,yen,gorengreinerjpg,gorengreinerprc,gorensteinplb,gorensteinjpg,yengorenprc}. This HRG model with excluded volume correction (EHRG) has been found to be in good agreement with LQCD up to temperature, T$\sim$ 0.140 GeV\cite{andronic,kapusta}.
 
 Since HRG (as well as EHRG) is a statistical model, the essential starting point is to find the partition function which in this case is just the partition function of an ideal gas summed over all the hadronic states and their resonances. While calculating the partition function at temperature (T), it is the zero temperature (and baryon chemical potential) hadron masses (M$_h$) that enters the Boltzmann factor, Exp(-M$_{h}/$T). It is well established fact that the chiral symmetry is an essential feature of QCD, the spontaneous breaking of which is responsible for the large part of  the quark mass called constituent quark mass, whence the hadrons. Further, LQCD as well as other effective model calculations at finite temperature shows that this symmetry is restored above so called chiral transition temperature (T$_{c}$) above which masses of approximate Goldstone modes drops down to current quark mass. Thus, since hadrons are made of quarks whose mass depends on temperature and chemical potential, it is T (and $\mu$) dependent hadron mass that should enter the partition function of HRG before computing any thermodynamical quantity. As we will see, taking into account this effect drastically changes the thermodynamics of hadronic matter at moderately high temperature. Since the HRG model has been used to calculate the transport properties as well, they are also non trivially affected.
 
 In this paper, we use SU(3) NJL model to compute masses of constituent quarks (u,d,s) as well as low lying mesons ($\pi,$ K, $\eta$ and $\eta'$). Since masses of the heavier mesons and baryons cannot be obtained in similar way, we use linear scaling rule in terms of constituent quarks for them. We compute all the thermodynamical quantities in EHRG at zero baryon chemical potential and confront them with the recent lattice QCD results\cite{prasad}. Further, we estimate shear viscosity coefficient within molecular kinetic theory generalized to gas of relativistic particles.
 Our basic motivation behind this work is to emphasize the fact that the hadron masses are temperature and density dependent which should be used to study the thermodynamics if the HRG model is used to estimate the same especially at temperatures where the effects of chiral symmetry become apparent. Further, although traditional HRG model is consistent with LQCD, it fails to include repulsive interactions which, we feel, are important too.  It is important to note that the way we have included T and $\mu$ dependent hadron masses is rather crude whence subjected to further improvement. 
 
 We organize the paper as follows.  In Sec. II we briefly describe the thermodynamically consistent excluded volume hadron resonance gas model. In Sec. III we present the results and make comment on the shear viscosity coefficient. Finally we summarize and conclude in Sec. IV.

\section{Hadron resonance gas model (HRG) with the medium modification of hadron masses}
The central quantity required to compute the thermodynamical quantities  is the partition function given as
\be
\log\mathcal{Z}(V,T,\mu)=\int dm(\rho_{b}(m)\log Z_{b}(V,m,T,\mu)+\rho_{f}(m)\log Z_{f}(V,m,T,\mu))
\label{pfhrg}
\ee
where the gas of hadrons is contained in volume V, at temperature T and baryon chemical potential $\mu$. $\rho_{b}$ and $\rho_{f}$ are the spectral densities of bosons and fermions respectively. Further, Z$_{b}$ and Z$_{f}$ are the partition functions of bosons and fermions respectively. The hadron properties enters into the model through spectral densities $\rho_{b/f}$. The (free) hadron resonance gas model is based on the result that the thermodynamics of interacting hadron resonance gas can be approximated by that of non interacting gas of hadrons provided all the resonances are included in the partition function\cite{Huovinen}. For such system the mass spectrum can be assumed to be sum over discrete hadronic states. 
\be
\rho_{b/f}=\sum_{a}^{M_{a}<\lambda}g_{a}\delta(m-M_{a})
\label{massspectrum}
\ee
 From the partition function defined by Eq. (\ref{pfhrg}) together with the mass spectrum given by Eq. (\ref{massspectrum}), the thermodynamical quantities can be readily obtained, $viz.$, Pressure $P(T,\mu)=T \lim_{V\rightarrow \infty}$ln$\mathcal{Z}(V,T,\mu)/V$, Baryon number density $n_{B}=\partial P(T,\mu)/\partial \mu$, entropy density $s(T,\mu)=\partial P(T,\mu)/\partial T$, energy density $\varepsilon(T,\mu)=Ts(T,\mu)-P(T,\mu)+\mu n_{B}(T,\mu)$, sound speed $C_{s}^{2}(T,\mu)=dP(T,\mu)/d\varepsilon(T,\mu)$.
 
 In the canonical non-interacting HRG model, one can take into account short range repulsive interactions via Van-der-Waals(VDW) correction in the volume of the system, i.e  by substitution V$\rightarrow$V$-v$N in the partition function given by Eq. (\ref{pfhrg}). Here, $v=4\frac{4}{3}\pi r_{h}^{3}$ is the proper volume parameter of the hadron with the hard core radius $r_{h}$\footnote{If we consider two hadrons with the hard
core radius $r_{h}$ just touching each other, they cannot come closer than the distance
$2r_{h}$ . Thus, for ”each pair” of hadrons the excluded volume is $v_{pair}=\frac{4}{3}\pi(2r_{h})^{3}$.
Whence for ”each hadron” the excluded volume will be half of this i.e $v=v_{pair} /2$
which turns out to be $4\frac{4}{3}\pi r_{h}^{3}$ .}. With this VDW corrected HRG model, i.e excluded volume hadron resonance gas model (EHRG), one obtains the transcendental equation for pressure as
 \be
 P^{EV}(T,\mu)=P_{ideal}(T,\tilde\mu)
 \ee
 where P$_{ideal}$ is the pressure computed within canonical HRG model and $\tilde\mu=\mu-vP^{EV}(T,\mu)$ is the effective chemical potential.
 The number density, energy density and entropy density can be readily obtained as

\be
n^{EV}(T,\mu)=\sum_{a}\frac{n^{id}_{a}(T,\tilde\mu)}{1+\sum_{a}v_{a}n_{a}^{id}(T,\tilde\mu)}
\ee
\be
\epsilon^{EV}(T,\mu)=\sum_{a}\frac{\epsilon^{id}_{a}(T,\tilde\mu)}{1+\sum_{a}v_{a}n_{a}^{id}(T,\tilde\mu)}
\ee
\be
s^{EV}(T,\mu)=\sum_{a}\frac{s^{id}_{a}(T,\tilde\mu)}{1+\sum_{a}v_{a}n_{a}^{id}(T,\tilde\mu)}
\ee
 
 In the temperature range where we are interested, the classical Boltzmann approximation is rather good approximation. In this approximation, VDW prescription is merely equivalent to the factor of Exp($-vP^{EV}/T$) to the ideal gas pressure. 
 
 We further extend excluded volume HRG model by taking into T and $\mu$ dependent hadron masses. To this end we use 2+1 flavor Nambu-Jona-Lasinio (NJL) model to obtain  T and $\mu$ dependent masses of light mesons ($\pi$,K, $\eta$, $\eta$')\cite{costa,hufner,torres}. For general reviews on NJL model, see\cite{klevansky, hatsuda, vogl}. We cannot treat heavier hadrons and resonances at par with light mesons to obtain their medium masses. So, for our purpose we use mass formulas bases on the work of Leupold\cite{leupold} and its generalization by Jankowski et al.\cite{jankowski}. Since the hadrons are made of either two or three quarks to make mesons and baryons respectively, we assume that the hadrons masses scale as
 \be
 M_{h}=(N_{q}-N_{s})\mathcal{M}_{q}+N_{s}\mathcal{M}_{s}+\kappa_{h}
 \label{scalingrule}
 \ee
where $N_{q}$ is the total number quarks in a given hadron and $N_{s}$ is the parameter which measures the strangeness content of the hadron. For the open strange hadrons, $N_{s}$ is simply number of strange ( antistrange) quarks. For hidden strange hadrons, $N_{s}=2/3$ for flavor singlet state while for flavor octet state, $N_{s}=4/3$. Further, $\mathcal{M}_{q}$ and $\mathcal{M}_{s}$ are the constituent masses of light (u,d) and strange (s) quarks respectively. $\kappa_{h}$ is the state dependent constant independent of current quark mass, $m_{q}$. In writing scaling rule (\ref{scalingrule})  we have assumed isospin symmetry i.e $m_{u}=m_{d}$. Eq. (\ref{scalingrule}) is used for all the hadrons except for light mesons. 

For the gas of hadrons at finite temperature and baryon chemical potential, we generalize the scaling rule (\ref{scalingrule}) as
\be
 M_{h}(T,\mu)=(N_{q}-N_{s})\mathcal{M}_{q}(T,\mu)+N_{s}\mathcal{M}_{s}(T,\mu)+\kappa_{h}
 \label{scalingrule2}
 \ee
We separate $M(T=0$, $\mu=0)$  part in above equation and absorb $\kappa_{h}$ in it to get
\be
 M_{h}(T,\mu)= M_{h}(T=0,\mu=0)+(N_{q}-N_{s})\mathcal{M}_{q}^{'}(T,\mu)+N_{s}\mathcal{M}_{s}^{'}(T,\mu)
 \label{scalingrule1}
 \ee
 where $\mathcal{M}_{q,s}^{'}$ is only medium dependent part of the constituent quark mass.
 
\section{Results and discussion}

\begin{figure}[h]
\vspace{-0.4cm}
\begin{center}
\begin{tabular}{c c}
 \includegraphics[width=8cm,height=8cm]{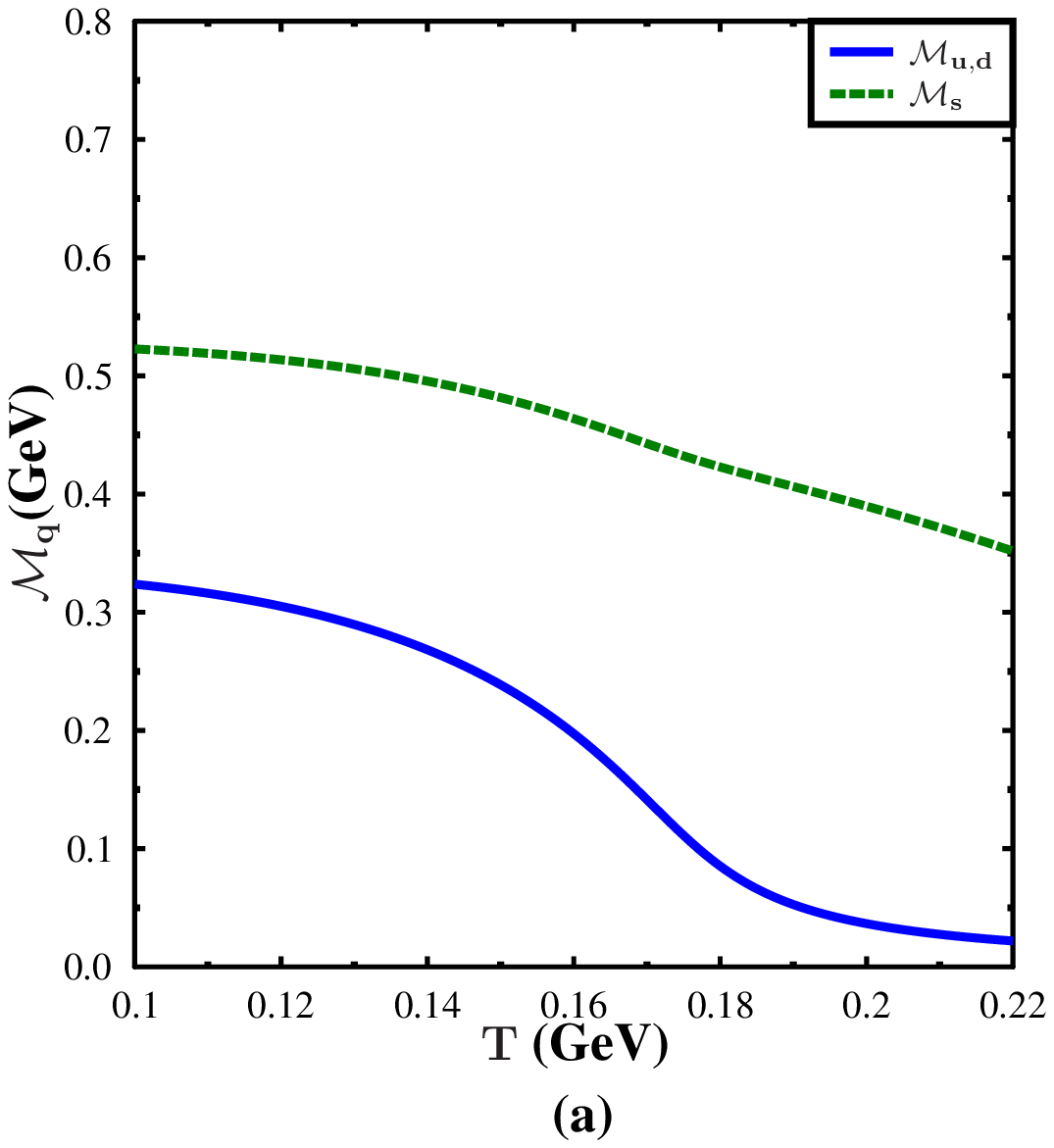}&
  \includegraphics[width=8cm,height=8cm]{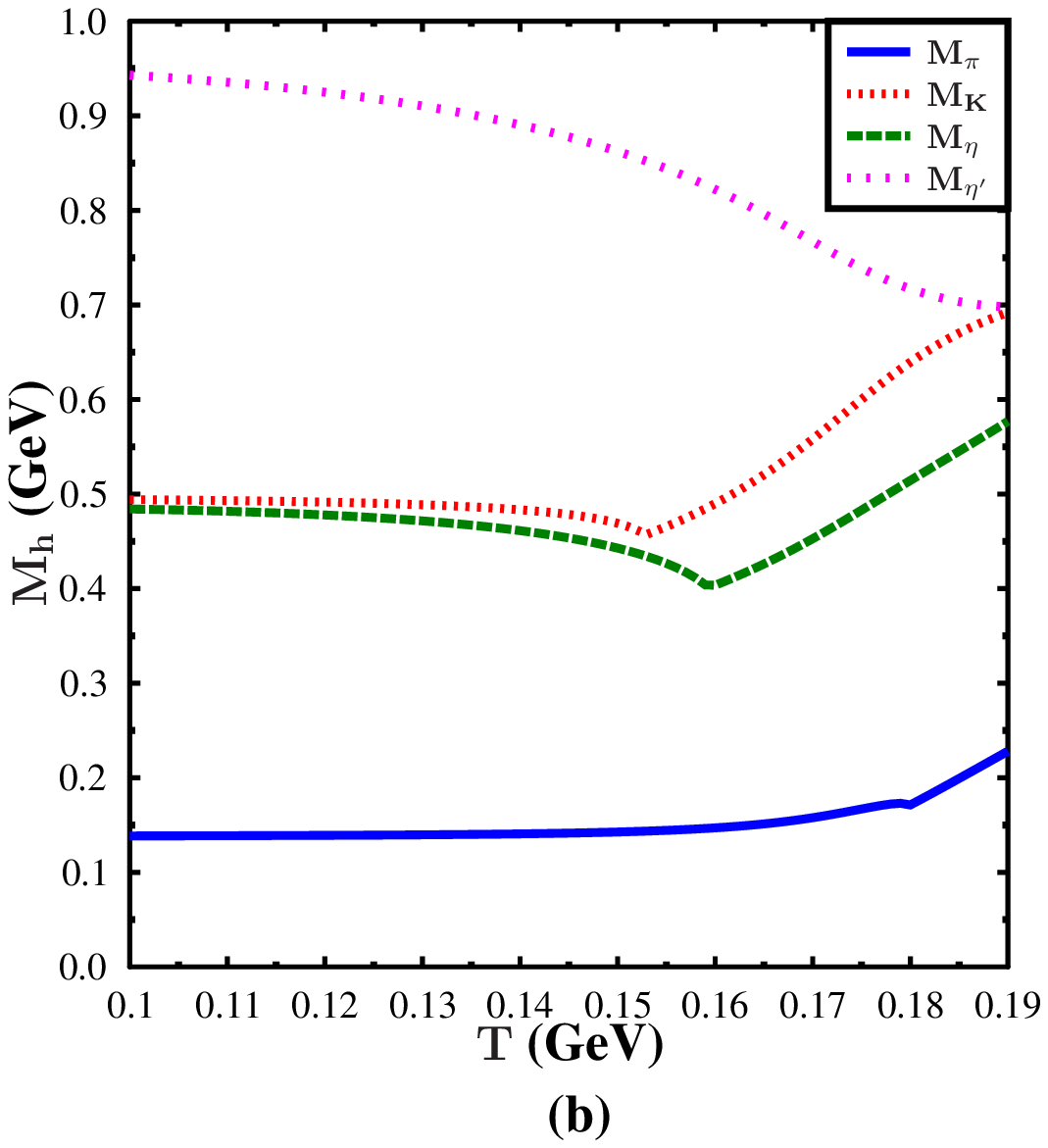}\\
 
  \end{tabular}
  \caption{(Color online)Left panel shows constituent quark masses as a function of temperature. Right panel shows light meson masses as a function of temperature.} 
\label{quark_meson_mass}
  \end{center}
 \end{figure}

\begin{figure}[h]
\vspace{-0.4cm}
\begin{center}
\begin{tabular}{c c c}
 \includegraphics[width=6cm,height=6cm]{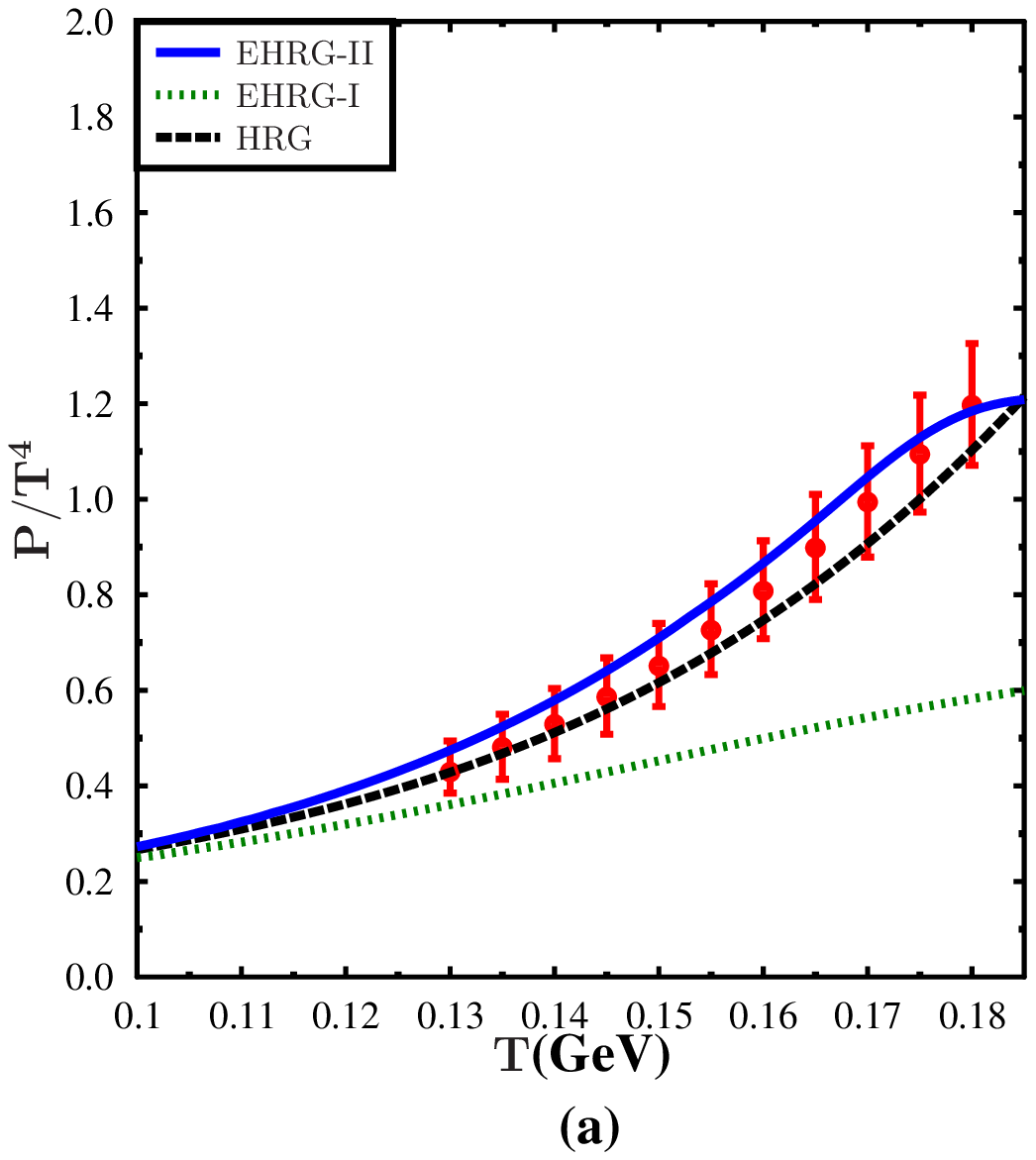}&
  \includegraphics[width=6cm,height=6cm]{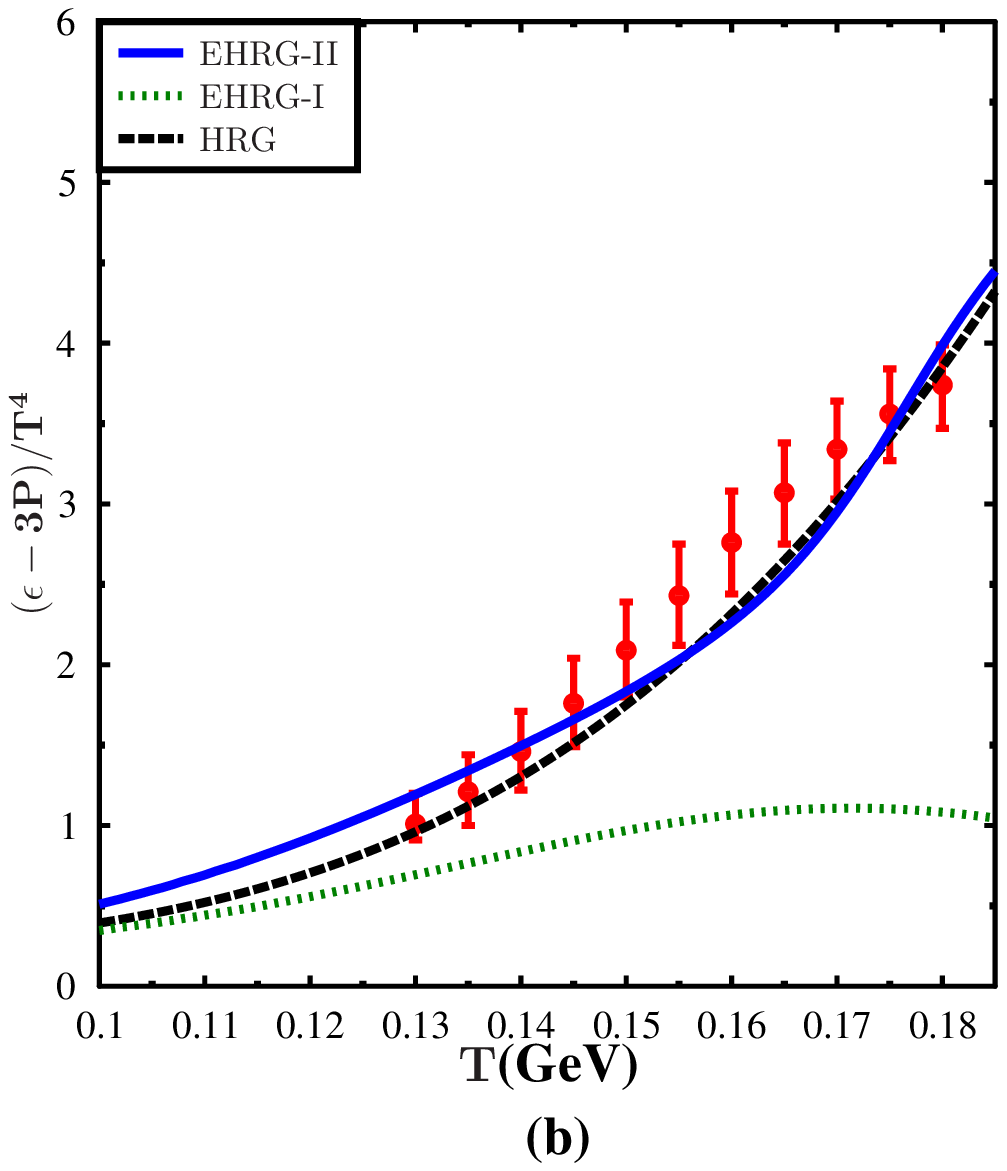}&
  \includegraphics[width=6cm,height=6cm]{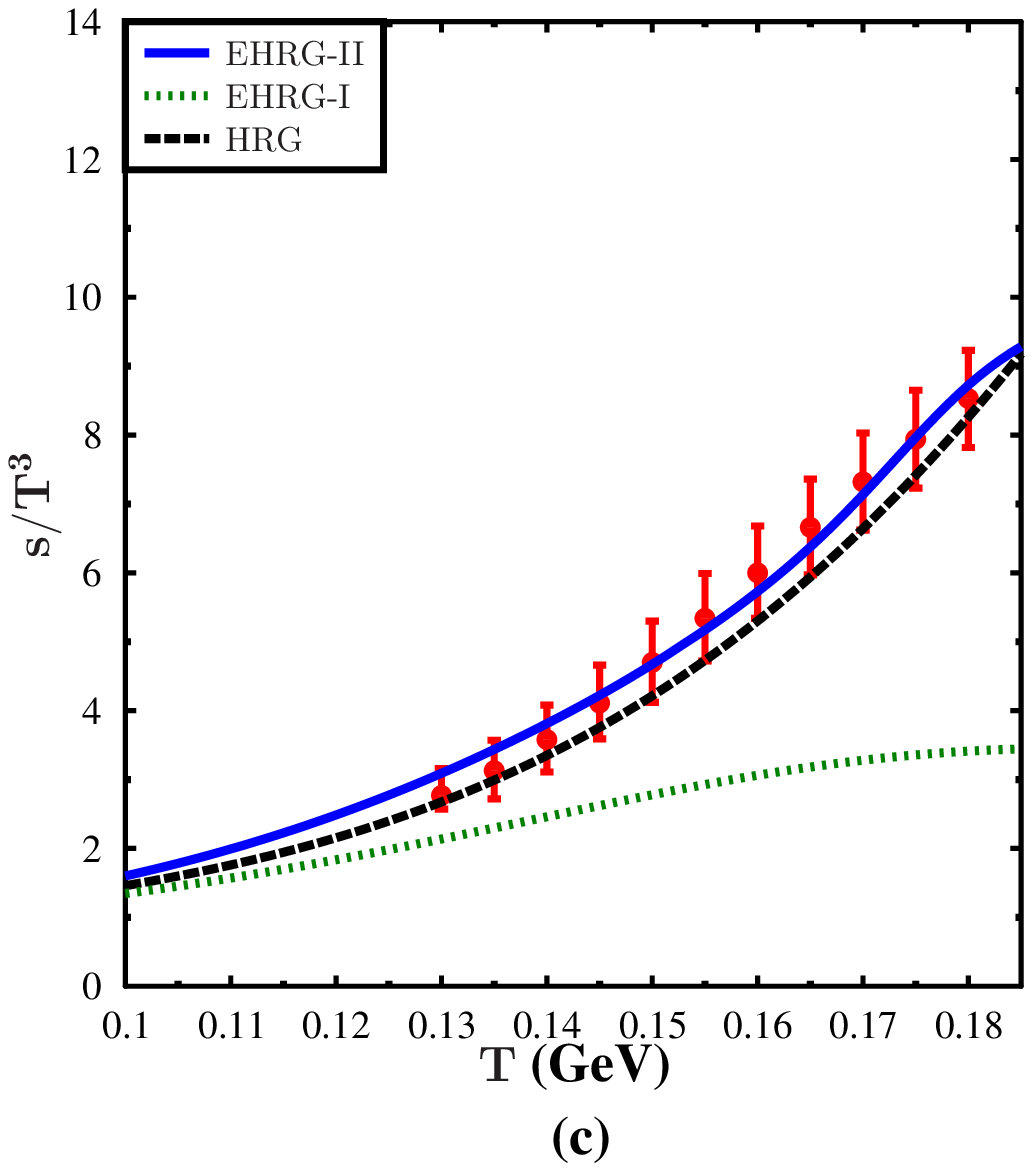}\\
  \end{tabular}
  \caption{(Color online)Results for thermodynamical quantities at $\mu=0$ GeV in EHRG-I and EHRG-II models with the mass dependent excluded volume parametrization. Black dashed curve corresponds to non-interacting hadron resonance gas model.  The lattice data (red circles) is from Ref. \cite{prasad}.} 
\label{thermo}
  \end{center}
 \end{figure}
 
   \begin{figure}[h]
\vspace{-0.4cm}
\begin{center}
 \includegraphics[width=9cm,height=9cm]{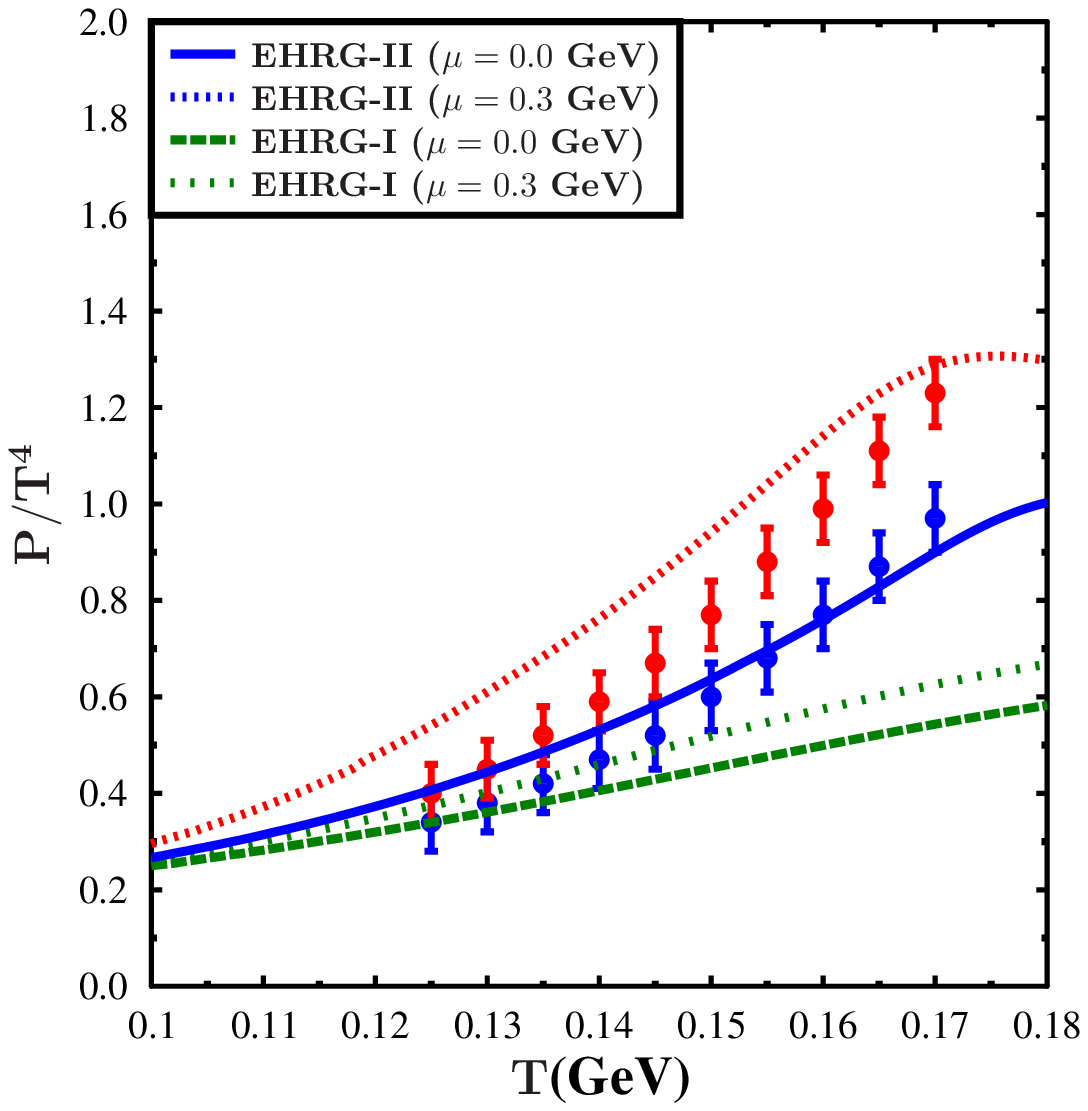}
  \caption{(Color online)Scaled pressure as a function of temperature at two different chemical potentials in EHRG-I and EHRG-II with mass dependent excluded volume parametrization and $\epsilon_{0}=0.7$ GeV fm$^{-3}$.} 
\label{pressure_mu}
  \end{center}
 \end{figure}

 We use the parameter set of Ref.\cite{hatsuda} to compute the masses of constituent quarks and light mesons using NJL model. To estimate the thermodynamic quantities using EHRG, we have taken all the hadrons and their resonances with the mass cutoff $\lambda=2.252$ GeV for baryons and 2.011 for mesons\cite{amseler}. The only unknown parameter which fixes the excluded volume HRG is the hardcore radius $r_{h}$ or the proper volume parameter $v$. It is customary in the literature to use uniform  values of hardcore radius for all the hadrons\cite{andronic,guruprc}. Baryonic hard core radius can be extracted from the short range repulsive interactions in nucleon-nucleon scattering processes. While it is legitimate to set hard core radius of all the baryons equal, detailed information regarding short range interaction between mesons is absent. Nevertheless, one can set same hard core radius to all mesons as that of baryons since meson charge radii are similar to the baryons\cite{heppe}.  But for our purpose we use the mass dependent hardcore radius as in Ref.\cite{kapusta}. In this scheme of parametrization, $v$ is chosen to be proportional to the mass of each hadron; $v=$M$/\epsilon_{0}$, $\epsilon_{0}$ is a constant which we fix to the value $0.9$ GeVfm$^{-3}$. We further generalize this scheme by taking into account T and $\mu$ dependent hadron masses.
 
 Fig. [\ref{quark_meson_mass}(a)] shows constituent quark mass as a function of temperature at $\mu=0$ GeV. We note that $\mathcal{M}_{u,d}$ decreases with with T and drops to current quark mass around 0.2 GeV. Albeit the strange quark mass ($\mathcal{M}_{s}$) also decreases with T, it does not attain its current quark mass. The temperature at which constituent quark masses attains its current quark mass is called transition temperature (T$_{c}$) corresponding to chiral restoration phase transition. [Fig. \ref{quark_meson_mass}(b)] shows low lying meson masses as a function of temperature at $\mu=0$ GeV.  The masses of (approximate) Goldstone modes show a weak dependence on the temperature as these are related to approximate SU(3) chiral symmetry breaking. In fact, one can show that for low temperatures dependence of mass of e.g. pion can be written as, using GOR relation at finite temperature\cite{GMORT}, 
 \be
 M_\pi^{2}(T ) = M_\pi^{2}(T=0)\bigg[1+\frac{1}{24}\frac{T^{2}} {f_\pi^{2}}\bigg]
 \ee
 Beyond the critical temperature the masses of pion and its chiral partner sigma become degenerate and increase with temperature. On the other hand, the $U_{A}(1)$ is not restored within the model and pion and eta masses are non degenerate. Further, cusps in Fig. [\ref{quark_meson_mass}(b)] corresponds to Mott transition\cite{hufner,costa,torres}.
 
 Results of the thermodynamical quantities are shown in Figs. (\ref{thermo}). We call EHRG without T, $\mu$ dependent hadron masses as EHRG-I and that with T, $\mu$ dependent hadron masses as EHRG-II. We note that the thermodynamical quantities computed within EHRG-II start deviating from EHRG-I at T$\sim 0.11$ GeV and this deviation is more pronounced above T$\sim 0.14$ GeV.  All the thermodynamical quantities are numerically larger in EHRG-II than in EHRG-I. This observation  can be explained by simply considering Boltzmann factor Exp(-M$_h(T,\mu)/T$). This factor is a measure of probability that the specific hadronic species of mass M is thermally excited at given temperature whence making a contribution to the thermodynamical quantities. Since masses of all the hadrons but pions, kaons and eta mesons decreases with temperature, they can be thermally excited abundantly with ease. As shown in Fig. \ref{quark_meson_mass}(b),  masses of the (approximately) Goldstone bosons do not change much around T$_{c}$, but the constituent quark masses do change significantly at this temperature  [Fig. \ref{quark_meson_mass}(a)]. In fact, it drops down to its current quark mass at T$_{c}$.  Since we expressed the heavy mesons and baryons masses in terms of constituent quarks [Eq. (\ref{scalingrule})] which contribute significantly at higher temperatures, we see the effect of T (and $\mu$) dependent hadron masses on the thermodynamics only at higher temperatures especially around transition temperature, while this effect is small at low temperatures where the pions and kaons are the dominating degrees of freedom.

  We note from Fig. \ref{thermo} that the thermodynamical quantities computed within conventional non-interacting hadron resonance gas (HRG) model better fits the lattice data than EHRG-I model. Such non-interacting HRG model corresponds to $v=0$. Thus, the lattice data seems to prefer zero excluded volume  parameter, whence the point particle picture of hadrons. But this observation does not invalidate EHRG-I model altogether. It has been shown in Ref.\cite{andronic} that non-interacting HRG model is problematic since the thermodynamical quantities rises very rapidly with temperature  and ultimately shows sign of Hagedorn divergence around T$_c$, while in EHRG-I thermodynamical quantities rise less steeply than that in free HRG. Further, the better agreement of HRG model over EHRG-I with the LQCD may be mere coincidence since we know from the experiments that nucleons, at least, are not point particles but they do have finite spatial extension\cite{heppe}. There is another experimental evidence that goes in favor of EHRG-I model.  The analysis of the data for particle number ratios of Au$+$Au (AGS)\footnote{Alternating Gradient Synchrotron} and Pb$+$Pb (SPS)\footnote{Super Proton Synchrotron} collisions suggest necessity to include excluded volume corrections in free HRG model\cite{yen}. Further, including the repulsive interactions via. excluded volume corrections in free HRG model and with the proper choice of excluded volume parameter, it is observed that EHRG-I agrees with lattice data up to T$\sim 0.14$ GeV\cite{andronic}. Eventually, by including  medium dependent hadron masses, the agreement of EHRG-I model with LQCD is observed at higher temperatures as may be noted from Fig. \ref{thermo}(a) where the pressure (normalized by T$^{4}$) computed with medium dependent hadron masses agrees with lattice data up to T$\sim 0.2$ GeV. Fig. (\ref{pressure_mu}) shows scaled pressure at finite chemical potential. We note that the pressure rises more rapidly even at finite $\mu$ in EHRG-II than in EHRG-I. We compare these results with LQCD simulations at finite chemical potential\cite{lqcdmu}. We note that at $\mu=0$, the estimations of EHRG-II are in better agreement with LQCD simulations that that of EHRG-I. Further, although the general behavior of the scaled pressure is similar we observe notable difference between LQCD simulations and EHRG-II estimations at $\mu=0.3$ GeV. But, it needs to be noted that these lattice simulations are estimated up to $\mathcal{O}(\mu^2)$.
 
 Fig. \ref{thermo}(b) shows trace anomaly (interaction measure) computed within EHRG-I and EHRG-II. We note that the trace anomaly rises rapidly in EHRG-II as compared to EHRG-I at high temperatures. Trace anomaly in EHRG-I shows decreasing behavior at high temperatures. The reason behind this is twofold. First, the suppression factor $\frac{1}{1+vn(T,\mu)}$ which decreases as temperature and chemical potential increases and hence all the thermodynamical quantities in EHRG-I are numerically smaller than that in ideal HRG.  Since there is no such suppression factor in HRG trace anomaly rises monotonically. Further, due to finite size of hadrons the pressure of hadron gas rises more rapidly as compare to energy density  whence the interaction measure decreases at high temperature in EHRG-I.  Strong suppression of thermodynamical quantities has also been observed earlier in Ref. \cite{vovchenko} where the authors studied EHRG with uniform hard core radius for all the hadrons. Although we have used different scheme of parametrization for hard core radius, suppression effect is still there. But in case of EHRG-II, since hard core radius is itself depend on temperature, the suppression effect is somehow diluted. Rapid rise of trace anomaly has also been observed in HRG model as well as extended HRG model which include continuum spectrum of hadrons (Hagedorn states) along with discrete spectrum\cite{greinerprc}.  
 
 Although our main purpose of this study is not to  fit the lattice data, this observation is rather crucial because as mentioned earlier, EHRG-I fails to explain the lattice data above T$\sim 0.14$ GeV\cite{andronic}. In our previous work in Ref. \cite{gurunpa},  we studied the extension of HRG model by including the Hagedorn density of states at finite temperature and density. We found rather good agreement with the lattice data of Ref.\cite{borsanyi} below T$=0.15$ GeV. In Ref. \cite{vovchenko} authors studied the two extensions of HRG model, $viz.$, HRG model with excluded volume effects (EHRG) and HRG with continuum mass spectrum (Hagedorn states) along with discrete mass spectrum of the hadrons. They observed that two models are not in agreement with the lattice QCD results of Ref. \cite{fodor} when considered separately. But when considered together, the suppression effects in EHRG and the enhancement effects due to Hagedorn states in HRG leads to better agreement with LQCD. From our observation that merely including the medium effects of hadrons in EHRG fit the LQCD quit well, it may be tempting to conclude that the effects of Hagedorn states can be alternatively simulated by including T and $\mu$ dependent hadron masses in EHRG.

\begin{figure}[h]
\vspace{-0.4cm}
\begin{center}
\begin{tabular}{c c}
 \includegraphics[width=8cm,height=8cm]{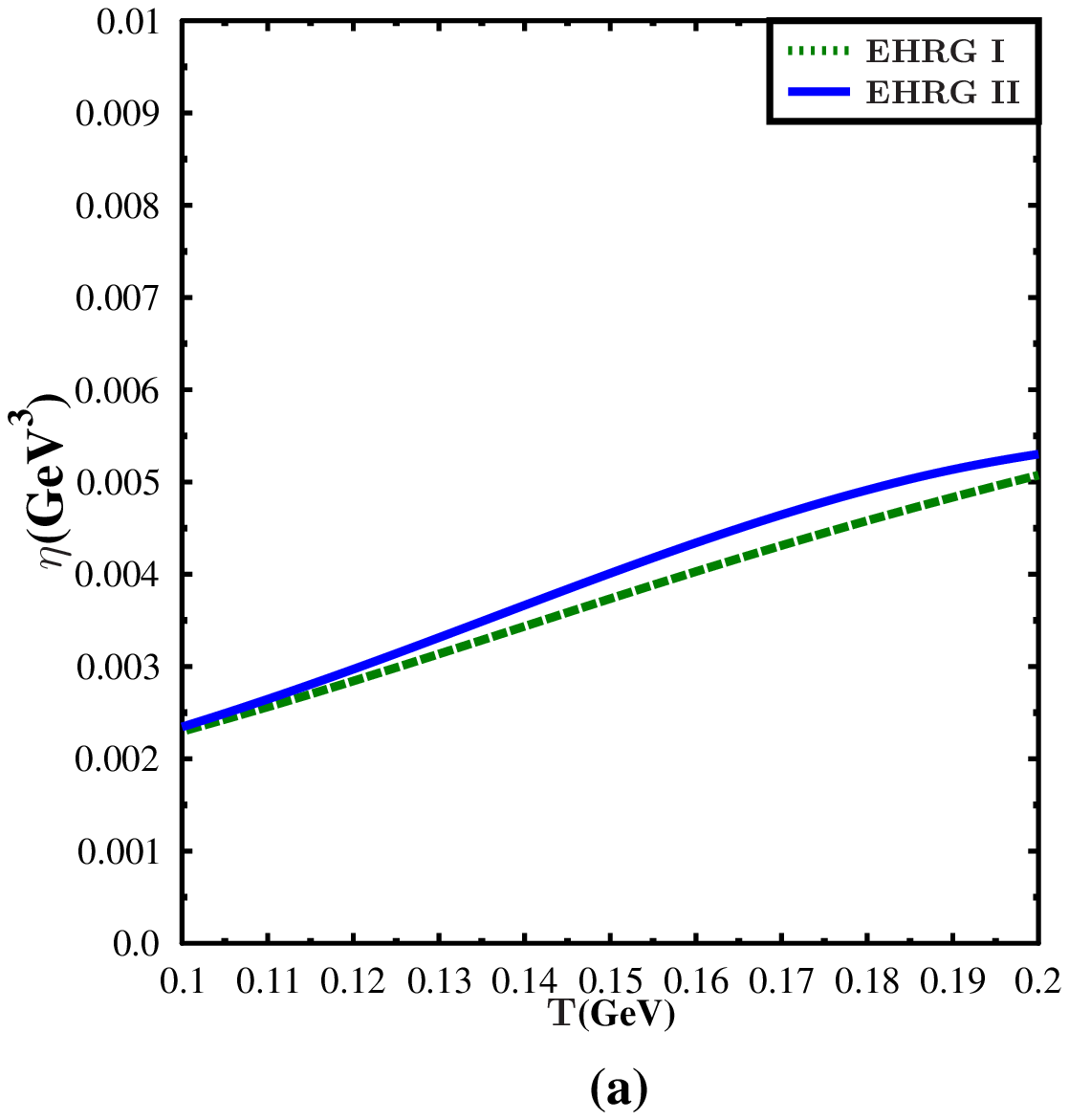}&
  \includegraphics[width=8cm,height=8cm]{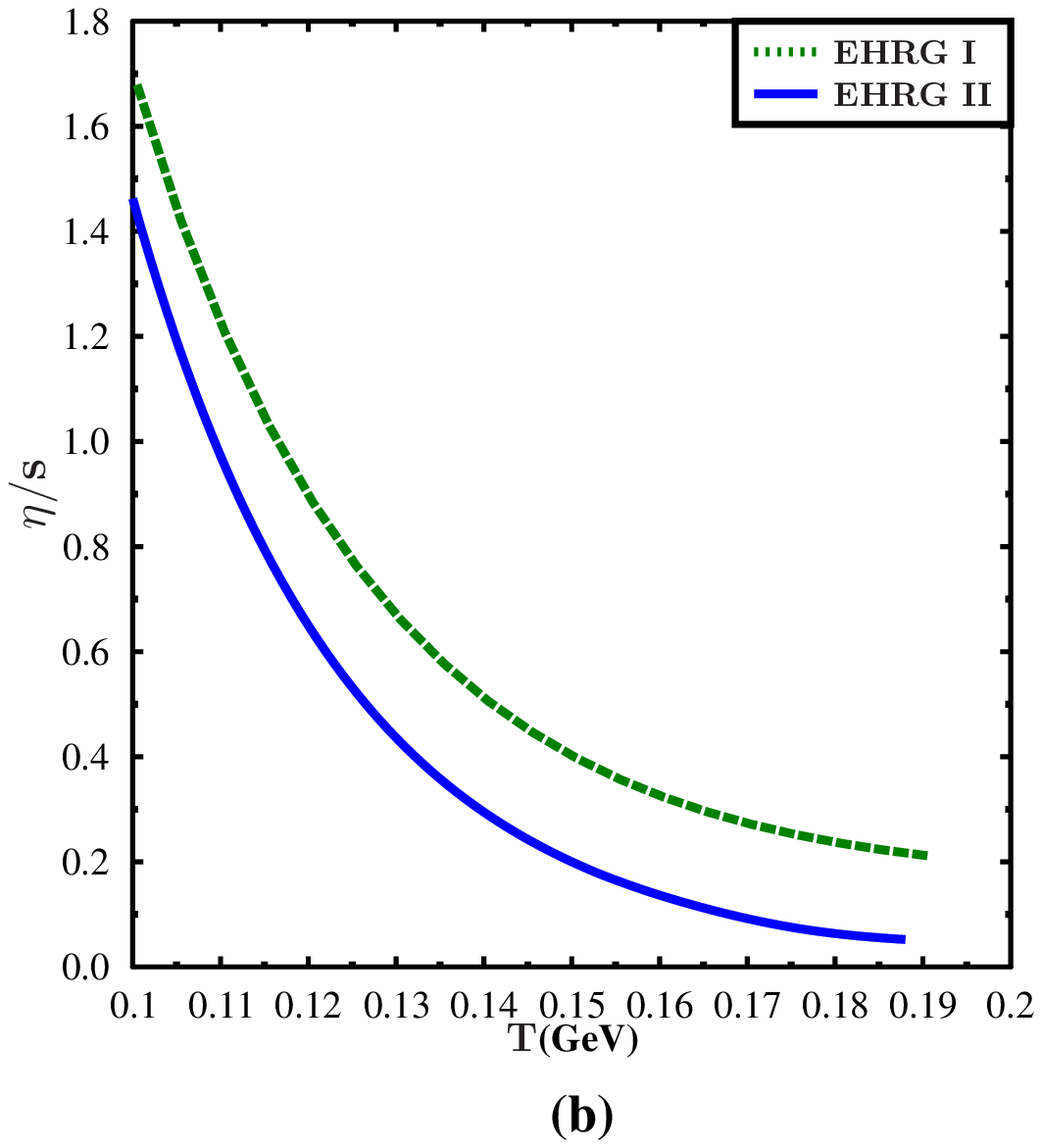}\\
  \end{tabular}
  \caption{(Color online)Left panel shows shear viscosity as a function of temperature in EHRG I and EHRG II. Right panel shows shear viscosity to entropy density in two models.} 
\label{etabis}
  \end{center}
 \end{figure}

The shear viscosity of the gas of relativistic particles of hard sphere radius $r_{h}$ can be calculated as\cite{gorenstein}
\be
\eta=\frac{5}{64\sqrt{8}r_{h}^{2}}\sum_{a}\frac{\langle\mid\vec{p}_{a}\mid\rangle n_{a}}{n}
\ee
 
where $\langle\mid\vec{p}_{i}\mid\rangle$ is the average thermal momentum of the ath species of hadron. $n_a$ is the number density of a$^{th}$ hadronic species and $n$ is the total number desnity with $n=\sum_{a}n_{a}$. Fig. \ref{etabis}(b) shows shear viscosity to entropy density ratio estimated in two models, EHRG-I and EHRG-II. We note that the effect T (and $\mu$) dependent hadron masses is also reflected in transport properties. Shear viscosity is propertional to average thermal momentum which is certainly affected by temperature dependent hadronic species in the system.  Nevertheless, the shear viscosity itself does not change much [Fig. \ref{etabis}(a)], but the ratio $\eta/s$ is smaller in EHRG-II model than in EHRG-I due to more rapid increase in the entropy density in former [Fig. \ref{thermo}(c)]. This effect is more important around transition temperature since the shear viscosity shows peculiar behavior around this temperature.  It may be interesting to compare these results with the results of Ref.\cite{greinerprl} where the authors estimated $\eta/s$ within EHRG model extended by inclusion of exponentially rising Hagedorn density of states. They observed that the inclusion of Hagedorn density of states significantly lowers $\eta/s$ and  this ratio approaches close to the KSS bound near T$_c$. Thus, they argued that the inclusion of Hagedorn states could explain the low value of shear viscosity  in the hadronic phase. Since we observed the same behavior of $\eta/s$ near T$_c$ but with the inclusion of medium dependent hadronic states, it may be again tempting to conclude that the effects of Hagedorn states can be alternatively simulated by including T and $\mu$ dependent hadron masses in EHRG.  
 
 \newpage
\section{Summary and Conclusion}
In present work we studied the effect of temperature (T) and baryon chemical potential ($\mu$) dependent hadron masses on the thermodynamics of hadronic matter. We used SU(3) Nambu-Jona-Lasinio model to compute T (and $\mu$) dependent masses of constituent quarks (u, d, s) as well as light mesons ($\pi,$ K, $\eta$ and $\eta'$ ). For heavier mesons and baryons we used linear scaling rule in terms of constituent quarks. We found that although the general behavior  don't change much,  taking into account T (and $\mu$) dependent hadron masses makes the thermodynamical quantities numerically larger than with zero temperature (and chemical potential) hadron masses at moderately high temperatures. This effect is more pronounced above T$=0.14$ GeV. We argued that this behavior can be explained by considering the Boltzmann factor Exp(-M$_{h}/T$) which is the measure of probability of thermal excitation of specific hadronic species at given temperature. Further, we found that including the medium effects of hadrons in EHRG explains LQCD data quite well over wide ranges of temperatures. Since it has been observed that the supression effect in EHRG together with the enhancement effect due to Hagedorn states leads to better agreement with LQCD, it may be tempting to conclude that the effects of Hagedorn states can be alternatively simulated by including T and $\mu$ dependent hadron masses in EHRG. At finite chemical potential, the EHRG-II is in better agreement with LQCD simulations than EHRG-I. We further emphasized the importance of taking into account this effect while estimating transport properties of hadronic matter especially around transition temperature.
 
 \def\karschkharzeev{F. Karsch, D. Kharzeev, and K. Tuchin, Phys. Lett. B
{\bf 663}, 217 (2008).}
\def\joglekar{J.C. Collins, A. Duncan, S.D. Joglekar, Phys. Rev. D {\bf 16}, 
438 (1977).}
\def\blaschke{J. Jankowski, D. Blaschke, M.Spalinski, Phys.Rev.D {\bf 87}, 105018
(2013). }
\def\gorenstein{M. Gorenstein, M. Hauer, O. Moroz, Phys.Rev.C {\bf 77}, 024911 (2008)}
\def\bugaev{K. Bugaev et al, Eur.Phys.J. A {\bf 49}, 30 (2013)}
\def\cpsingh{S.K. Tiwari, P.K. Srivastava, C.P. Singh, Phys.Rev. C {\bf 85},
014908 (2012).}
\def\chen{J.-W. Chen, Y-H. Li, Y.-F. Liu, and E. Nakano, Phys. Rev. D {\bf 76},
114011 (2007)}
\def\chennakano{J.-W. Chen, and E. Nakano, Phys. Lett. B {\bf 647}, 371 (2007)}
\def\itakura{K. Itakura, O. Morimatsu, and H. Otomo, Phys. Rev. D {\bf 77}, 014014
(2008)}
\def\cleymans{J. Cleymans, H. Oeschler, K. Redlich, and S. Wheaton, Phys.
Rev. C {\bf 73}, 034905 (2006)}
\def\hirano{P. Romatschke and U. Romatschke, Phys. Rev. Lett. {\bf 99}, 172301 (2007); T. Hirano and M. Gyulassy, Nucl. Phys. {\bf A 769}, 71, (2006).} 
\def\kss{P. Kovtun, D.T. Son and A.O. Starinets, Phys. Rev. Lett. {\bf 94},
 111601 (2005).}
\def\sasakiqp{C. Sasaki and K.Redlich,{\PRC{79}{055207}{2009}}.}
\def\sasakinjl{C. Sasaki and K.Redlich,{\NPA{832}{62}{2010}}.}
\def\ellislet{I.A. Shushpanov, J. Kapusta and P.J. Ellis,{\PRC{59}{2931}{1999}}
; P.J. Ellis, J.I. Kapusta, H.-B. Tang,{\PLB{443}{63}{1998}}.}
\def\prakashwiranata{A. Wiranata and Madappa Prakash, Phys. Rev. C {\bf 85}, 054908 (2012).}
\def\purnendu{P. Chakraborty and J.I. Kapusta {\PRC{83}{014906}{2011}}.}
\def\greco{S.Plumari,A. Paglisi,F. Scardina and V. Greco,{\PRC{83}{054902}{2012}.}}
\def\bes{H. Caines, arXiv:0906.0305 [nucl-ex], 2009.}
\def\greinerprl{J. Noronha-Hostler,J. Noronha and C. Greiner, 
{\PRL{103}{172302}{2009}}.}
\def\greinerprc{J. Noronha-Hostler,J. Noronha and C. Greiner
, {\PRC{86}{024913}{2012}}.}
\def\igorgreiner{J. Noronha-Hostler, C. Greiner and I. Shovkovy,
, {\PRL{100}{252301}{2008}}.}
\def\nakano{J.W. Chen,Y.H. Li, Y.F. Liu and E. Nakano,
 {\PRD{76}{114011}{2007}}.}
\def\itakura{K. Itakura, O. Morimatsu, H. Otomo, {\PRD{77}{014014}{2008}}.}
\def\wang{M.Wang,Y. Jiang, B. Wang, W. Sun and H. Zong, Mod. Phys. lett.
{\bf A76}, 1797,(2011).}
\def\rischkegorenstein{.D.H. Rischke, M.I. Gorenstein, H. Stoecker and
W. Greiner, Z.Phys. C {\bf 51}, 485 (1991).}
\def\hmnjl{Amruta Mishra and Hiranmaya Mishra, {\PRD{74}{054024}{2006}}.}
\def\pdgb{C. Amseler {\it et al}, {\PLB{667}{1}{2008}}.}
\def\shuryak{E.V. Shuryak, Yad. Fiz. {\bf 16},395, (1972).}
\def\leupold{S. Leupold, J. Phys. G{\bf32},2199,(2006)}
\def\peter{A. Andronic, P. Braun-Munzinger , J. Stachel and M. Winn,
{\PLB{718}{80}{2012}}}
\def\blum{M. Blum, B. Kamfer, R. Schluze, D. Seipt and U. Heinz,{\PRC{76}{034901}{2007}}.}
\def\jaminplb{M. Jamin{\PLB{538}{71}{2002}}.}
\def\ghosh{Sabyasachi Ghosh{\PRC{90}{025202}{2014}}.}
\def\csernai{L.P. Csernai, J.I. Kapusta and L.D. McLerran,{\PRL{97}{152303}{2006}}.}
\def\hagedorn{R. Hagedorn, Nuovo Cim. Suppl. 3,147 (1965); Nuovo Sim. A56,1027 (1968).}
\def\torieri{G. Torrieri and I. Mishustin,{\PRC{77}{034903}{2008}}.}
\def\fernandez{D. Fernandiz-Fraile and A.G. Nicola,{\PRL{102}{121601}{2009}}.}
\def\caron{S.Caron,{\PRD{79}{125009}{2009}}.}
\def\latticemeyer{H.B. Meyer,{\PRL{100}{162001}{2008}}.} 
\def\romatschke{P.Romatscke and D.T. Son,{\PRD{80}{065021}{2009}}.}
\def\moore{G.D. Mooore and O. Sarem, J. High Energy Phys. JHEP0809(2008)015.}
\def\dobado{A.Dobado and J. M. Torres-Rincon {\PRD{86}{074021}{2012}}.}
\def\daniel{D. Fernandez-Fraile, {\PRD{83}{065001}{2011}}.}
\def\gurunpa{G.P. Kadam, H. Mishra, Nuclear Physics A {\bf 934}, 133 (2015).}
\def\gurumpla{G.P. Kadam, Mod.Phys.Lett. A {\bf 30},  1550031 (2015).}
\def\demir{N. Demir, S.A. Bass, Phys. Rev. Lett. {\bf 102}, (2009) 172302.}
\def\phsd{V. Ozvenchuk, O. Linnyk, M.I. Gorenstein, E.L. Bratkovskaya, W. Cassing, Phys. Rev. C {\bf 87}, (2013) 064903.}
\def\ehrgrishke{D. H. Rischke, M. I. Gorenstein, H. Sto ̈cker, and W. Greiner,
Z. Phys. C {\bf 51}, 485 (1991).͒}
\def\ehrgclaymans{J. Cleymans, M. I. Gorenstein, J. Stalnacke, and E. Suhonen,
Phys. Scr. {\bf 48}, 277 (1993͒).}
\def\gavin{S. Gavin,  Nucl.Phys. A {\bf 435}, 826 (1985).}
\def\cannoni{Mirco Cannoni, Phys. Rev. D {\bf 89}, 103533 (2014).}
\def\gelmini{P. Gondolo and G. Gelmini, Nucl. Phys. B {\bf 360}, 145 (1991).}
\def\jeon{G.S. Denicol, C. Gale,  S. Jeon,  and J. Noronha, Phys. Rev. C {\bf 88}, 064901 (2013).}
\def\sinha{A. Buchel, R. C. Myers and A. Sinha, JHEP {\bf 0903}, 084 (2009).}
\def\heppe{P. Braun-Munzinger, I. Heppe, J. Stachel, Phys. Lett. B {\bf 465}, 15 (1999).}
\def\bugaev{K.A. Bugaev, D.R. Oliinychenko, A.S. Sorin, and G.M. Zinovjev, arxive:1208.5968v1.}
\def\cleymans{J. Cleymans, H. Oeschler, K. Redlich, S. Wheaton, Phys. Rev. C {\bf 73}, 034905 (2006).}
\def\moore{E. Lu, G. D. Moore, Phys. Rev. C {\bf 83}, 044901 (2011).}
\def\amseler{C. Amseler, et al., Phys. Lett. B {\bf 667}, 1 (2008).}
\def\moroz{O. Moroz, Ukr.J.Phys. {\bf 58}, 1127 (2013).}
\def\gurunxt{Guru Prakash Kadam, H. Mishra, In preparation.}
\def\yen{G. D. Yen, M. I. Gorenstein, W. Greiner, and S. N. Yang, Phys.Rev. C {\bf 56}, 2210 (1997).}
\def\sarkarghosh{ S. Ghosh, G. Krein, S. Sarkar, Phys.Rev. C {\bf 89}, 045201 (2014).}
\def\finazzo{ S. I. Finazzo, R. Rougemont, H. Marrochio, J. Noronha, JHEP {\bf 1502}, 051 (2015).}
\def\khvorostukhin{A.S. Khvorostukhin, V.D. Toneev, D.N. Voskresensky, Nucl. Phys. A {\bf 845}, 106 (2010).}
\def\sghosh{S. Ghosh, Int. J. Mod. Phys. A {\bf 29}, 1450054 (2014).}
\def\sghoshnucl{S. Ghosh, Phys. Rev. C {\bf 90}, 025202 (2014).}
\def\sghoshnjl{S. Ghosh,  A. Lahiri, S. Majumder, R. Ray, S. K. Ghosh, Phys. Rev. C {\bf 88}, 068201 (2013).}
\def\cohen{ T.D. Cohen, Phys. Rev. Lett. {\bf 99}, 021602 (2007).}
\def\rebhan{ A. Rebhan and D. Steineder, Phys. Rev. Lett. {\bf 108}, 021601 (2012).}
\def\mamo{ K.A. Mamo, JHEP {\bf 70}, 1210 (2012).}
\def\weise{R. Lang, N. Kaiser, and W. Weise, arxive:1506.02459}
\def\dobadojuan{A. Dobado, F. J. Llanes-Estrada and J. M. Torres-Rincon, Phys. Lett. B {\bf 702}, 43 (2011).}
\def\gyulassy{P. Danielewicz and M. Gyulassy, Phys.Rev. D {\bf 31}, 53 (1985).}
\def\borsanyi{S. Borsányi, et al., JHEP {\bf 1011}, 077 (2010).}
\def\prasad{Bazavov et al., Phys. Rev. D {\bf 90}, 094503 (2014).}
\def\WPblk{A. Wiranata and M. Prakash, Nucl. Phys. A {\bf 830}, 219–222 (2009).}
\def\prakash{M. Prakash, M. Prakash, R. Venugopalan and G. Welke, Phys.Rept. {\bf 227}, 321-366 (1993).}
\def\wiranatakoch{A. Wiranata, V. Koch and  M. Prakash, X.N. Wang, J.Phys.Conf.Ser. {\bf 509}, 012049 (2014).}
\def\wiranataprc{A. Wiranata and M. Prakash, Phys.Rev. C {\bf 85}, 054908 (2012).}
\def\wiraprapurn{ A. Wiranata, M. Prakash and  P. Chakraborty, Central Eur.J.Phys. {\bf 10}, 1349-1351 (2012).}
\def\wiranatademir{N. Demir and A. Wiranata, J.Phys.Conf.Ser. {\bf 535}, 012018 (2014).}
\def\gorensteinplb{M. Gorenstein, V. Petrov and G. Zinovjev, Phys.Lett. B {\bf 106}, 327-330 (1981).}
\def\goregranddonprc{G. D. Yen, M. Gorenstein, W. Greiner and Shin Nan Yang, Phys. Rev. C {\bf 56}, 2210 (1997).}
\def\gorensteinjpg{M. Gorenstein, H. St\"{o}ker, G. D. Yen, Shin Nan Yang and W. Greiner, J. Phys. G {\bf 24}, 1777 (1998).}
\def\yengorenprc{G.D. Yen and M. Gorenstein, Phys. Rev. C {\bf 59}, 2788 (2009).}
\def\gorengreinerjpg{M. Gorenstein, W. Greiner and Shin Nan Yang, J. Phys. G {\bf 24}, 725 (1998).}
\def\gorengreinerprc{M. Gorenstein, M. Ga'zdicki and W. Greiner, Phys. Rev. C {\bf 72}, 024909 (2005).}
\def\tawfik{A. Tawfik and M. Wahba, Ann. Phys. {\bf 522}, 849-856 (2010).}
\def\jeonyaffe{S. Jeon and L. Yaffe, Phys.Rev. D {\bf 53}, 5799-5809 (1996).}
\def\liao{J. Liao, V. Koch, Phys. Rev. D {\bf 81}, 014902 (2010).}
\def\Khov{A. Khvorostukhin, V. Toneev, D. Voskresensky, Nucl. Phys. A {\bf 845}, 106–146, (2010).}
\def\nicola{D. Fernandez-Fraile, A. Gomez-Nicola, Eur.Phys.J. C {\bf 62}, 37-54 (2009).}
\def\Huovinen{P. Huovinen, P. Petrecky, Nucl. Phys. A {\bf 837}, 26 (2010).}
\def\borsonyi{S. Borsányi, et al., JHEP {\bf 1009}, 073 (2010).}
\def\GMOR{J. Gasser and H. Leutwyler, Nucl. Phys. B{bf\ 250}, 465 (1985).}
\def\Huovinen{P. Huovinen, P. Petrecky, Nucl. Phys. A {\bf 837}, 26 (2010).}
\def\welke{G. M. Welke, R. Venugopalan, M. Prakash, Phys. Lett. B {\bf 245}, 137 (1990).}
\def\prakashvenu{R. Venugopalan, M. Prakash, Nucl. Phys. A {\bf 546}, 718 (1992).}
\def\andronic{A. Andronic, P. Braun-Munzinger, J. Stachel, M. Winn, Phys. Lett. B {\bf 718}, 80 (2012).}
\def\kapusta{M. Albright, J. Kapusta, and C. Young, Phys. Rev. C {\bf 90}, 024915 (2014).}
\def\bhattacharyya{Abhijit Bhattacharyya, Rajarshi Ray, Subhasis Samanta, and Subrata Sur, Phys. Rev. C {\bf 91}, 041901 (2015).}
\def\garg{ P. Garg,  D.K. Mishra, P.K. Netrakanti, B. Mohanty, A.K. Mohanty, B.K. Singh, N. Xu, Phys.Lett. B {\bf 726}, 691 (2013).}
\def\guruprc{G.P. Kadam, H. Mishra, Phys. Rev.C {\bf 92}, 035203 (2015).}
\def\costa{P. Costa, M.C. Ruivo, C.A. de Sousa, Phys. Rev. C {\bf 70}, 025204 (2004).}
\def\klevansky{S. P. Klevansky, Rev. Mod. Phys. {\bf 64}, 649 (1992).}
\def\hatsuda{T. Hatsuda and T. Kunihiro, Phys. Rep. {\bf 247}, 221 (1994).}
\def\vogl{U. Vogl and W. Weise, Prog. Part. Nucl. Phys. {\bf 27}, 195 (1991).}
\def\hufner{P. Rehberg, S. P. Klevansky, and J. H\"{u}fner, Phys. Rev. C {\bf 53}, 410 (1996).}
\def\leupold{S. Leupold, J. Phys. G {\bf 32}, 2199 (2006).}
\def\jankowski{J. Jankowski, D. Blaschke, and M. Spali\'{n}ski, Phys. Rev. D {\bf 87}, 105018 (2013).}
\def\vovchenko{V. Vovchenko, D. V. Anchishkin and M. I. Gorenstein, Phys. Rev. C {\bf 91}, 024905 (2015)}
\def\fodor{S. Borsanyi, Z. Fodor, C. Hoelbling, S. D. Katz, S. Krieg, and K. K. Szabo, Phys. Lett. B {\bf 730}, 99 (2014).}
\def\cleymansatz{J. Cleymans and H. Satz, Z. Phys. C {\bf 57}, 135 (1993).}
\def\magestro{P. Braun-Munzinger, D. Magestro, K. Redlich, and J. Stachel,
Phys. Lett. B {\bf 518}, 41 (2001).}
\def\rafelski{J. Rafelski and J. Letessier, Nucl. Phys. A {\bf 715}, 98 (2003).}
\def\stachelandronic{A. Andronic, P. Braun-Munzinger, and J. Stachel, Nucl. Phys.
A 772, 167 (2006).}
\def\Becattini{F. Becattini, J. Cleymans, A. Keranen, E. Suhonen, and
K. Redlich, Phys. Rev. C 64, 024901 (2001).}
\def\torres{J.M. Torres-Rincon, B. Sintes and J. Aichelin, Phys. Rev. C {\bf 91}, 065206 (2015).}
\def\bohr{A. Bohr, B. Mottelson, Nuclear Structure, vol. 1, Benjamin, New York, 1969, p. 251.}
\def\GMORT{J. Gasser and H. Leutwyler, Phys. Lett. B {\bf 184}, 83 (1987).}
\def\lqcdmu{S. Borsonyi, et al., JHEP {\bf 1208}, 053 (2012).}


\begin{thebibliography}{99}

\bibitem{Huovinen}\Huovinen
\bibitem{welke}\welke
\bibitem{prakashvenu}\prakashvenu
\bibitem{prasad}\prasad
\bibitem{cleymansatz}\cleymansatz
\bibitem{Becattini}\Becattini
\bibitem{magestro}\magestro
\bibitem{rafelski}\rafelski
\bibitem{prakash}\prakash
\bibitem{prakashwiranata}\prakashwiranata
\bibitem{khvorostukhin}\khvorostukhin
\bibitem{gorenstein}\gorenstein
\bibitem{greinerprc}\greinerprc
\bibitem{cpsingh}\cpsingh
\bibitem{tawfik}\tawfik
\bibitem{greinerprl}\greinerprl
\bibitem{gurunpa}\gurunpa
\bibitem{gurumpla}\gurumpla
\bibitem{guruprc}\guruprc
\bibitem{bhattacharyya}\bhattacharyya
\bibitem{garg}\garg
\bibitem{heppe}\heppe
\bibitem{bohr}\bohr
\bibitem{ehrgrishke}\ehrgrishke
\bibitem{ehrgclaymans}\ehrgclaymans
\bibitem{yen}\yen
\bibitem{gorengreinerjpg}\gorengreinerjpg
\bibitem{gorengreinerprc}\gorengreinerprc
\bibitem{gorensteinplb}\gorensteinplb
\bibitem{gorensteinjpg}\gorensteinjpg
\bibitem{yengorenprc}\yengorenprc
\bibitem{andronic}\andronic

\bibitem{kapusta}\kapusta
\bibitem{costa}\costa
\bibitem{torres}\torres
\bibitem{klevansky}\klevansky
\bibitem{hatsuda}\hatsuda
\bibitem{vogl}\vogl
\bibitem{hufner}\hufner
\bibitem{leupold}\leupold
\bibitem{jankowski}\jankowski
\bibitem{amseler}\amseler
\bibitem{GMORT}\GMORT
\bibitem{lqcdmu}\lqcdmu
\bibitem{vovchenko}\vovchenko
\bibitem{borsanyi}\borsanyi
\bibitem{fodor}\fodor


\vfil
\end{thebibliography}
\end{document}